\documentclass[aps,prl,preprint,groupedaddress,showpacs,amsmath,amssymb,floatfix]{revtex4-1}
\usepackage{graphicx}
\usepackage{bm}
\usepackage{color}
\usepackage[normalem]{ulem}
\usepackage{cleveref}
\usepackage{amsmath}

\usepackage{amssymb}
\usepackage{epsfig}
\usepackage{epstopdf}
\usepackage{gensymb}
\usepackage[utf8x]{inputenc}
\usepackage{comment}
\usepackage{ulem}

\begin{document}

\title{Quantum Criticality in Twisted Transition Metal Dichalcogenides}
\author{Augusto Ghiotto$^{1}$}
\author{En-Min Shih$^{1}$}
\author{Giancarlo S. S. G. Pereira$^{1}$}
\author{Daniel A. Rhodes$^{2}$}
\author{Bumho Kim$^{2}$}
\author{Jiawei Zang$^{1}$}
\author{Andrew J. Millis$^{1,3}$}
\author{Kenji Watanabe$^{4}$}
\author{Takashi Taniguchi$^{4}$}
\author{James C. Hone$^{2}$}
\author{Lei Wang$^{1,5 \ast}$}
\author{Cory R. Dean$^{1,\ast}$}
\author{Abhay N. Pasupathy$^{1,6 \ast}$}

\affiliation{$^{1}$Department of Physics, Columbia University, New York, NY 10027, USA}
\affiliation{$^{2}$Department of Mechanical Engineering, Columbia University, New York, NY 10027, USA}
\affiliation{$^{3}$Center for Computational Quantum Physics, Flatiron Institute, New York, NY 10010 USA}
\affiliation{$^{4}$National Institute for Materials Science, Namiki 1-1, Tsukuba, Ibaraki 305-0044, Japan}
\affiliation{$^{5}$National Laboratory of Solid-State Microstructures, School of Physics, Nanjing University, Nanjing, 210093, China}
\affiliation{$^{6}$Condensed Matter Physics and Materials Science Division, Brookhaven National Laboratory, Upton, New York 11973, USA},
\affiliation{$^{\ast}$Corresponding authors, Email: leiwang@nju.edu.cn,  cd2478@columbia.edu, apn2108@columbia.edu}


\maketitle

\textbf{
In moiré heterostructures, gate-tunable insulating phases driven by electronic correlations have been recently discovered. Here, we use transport measurements to characterize the gate-driven metal-insulator transitions and the metallic phase in twisted WSe$_2$ near half filling of the first moiré subband. We find that the metal-insulator transition as a function of both density and displacement field is continuous. At the metal-insulator boundary, the  resistivity displays strange metal behaviour at low temperature with dissipation comparable to the Planckian limit. Further into the metallic phase, Fermi-liquid behaviour is recovered at low temperature which evolves into a quantum critical fan at intermediate temperatures before eventually reaching an anomalous saturated regime near room temperature. An analysis of the residual resistivity indicates the presence of strong quantum fluctuations in the insulating phase. These results establish twisted WSe$_2$ as a new platform to study doping and bandwidth controlled metal-insulator quantum phase transitions on the triangular lattice.}

In strongly correlated materials such as cuprates, pnictides and heavy Fermion systems, a plethora of ordered phases have been observed, including Mott insulators, superconductors and density waves. The most intriguing electronic properties of these materials are not found in these ordered phases, but instead in the metallic phases that are adjacent to them. Near criticality, these metallic phases exhibit anomalous transport properties that defy description by the Landau Fermi liquid paradigm \cite{Varma, Qimiao, Zaanen, Imada}. One striking manifestation of this is the temperature and magnetic field dependence of the resistance of the metallic phase, which deviates strongly from the expected $T^2$ or $B^2$ dependence predicted by Fermi liquid theory. Other transport coefficients, including the Hall coefficient, Nernst effect and thermal conductivity, display anomalous properties as well \cite{Taillefer}. These "strange metal" properties are often associated with a quantum critical point, a second order quantum phase transition driven by a control parameter such as doping, applied field or pressure \cite{Qimiao, Varma}. Understanding the nature of quantum fluctuations and how they give rise to anomalous metallic properties is among the most important open problems in condensed matter physics.

Moiré materials exhibit a number of quantum phases \cite{Cao2018, Cao2018a, Yankowitz2019, Burg2019, Liu, Shen2020, Cao_double, Wang2020, Regan2020, Tang2020, Chen, Shi, Chen2019a, Hao, Park, Kerelsky, Cao_nem, Rubio-Verdu}, including insulators at integer and fractional fillings of the moiré lattice. The importance of strong electron correlations to the insulating phases at integer filling of the moiré lattice has been unambiguously established, but much less is known either experimentally or theoretically about the nature of the metallic state that exists nearby or the transitions that connect metallic and insulating states. Metal-insulator transitions can generally be driven by either carrier doping or by tuning electronic bandwidth. In most materials, these transitions are first order due to electron-lattice coupling, with additional complications introduced by chemical doping disorder. In this work, we present twisted homobilayer WSe$_2$ (tWSe$_2$) as an ideal platform to realize continuous metal-insulator transitions. In this system, both the electron density and the electronic structure can be tuned in a facile manner using electrostatic gates without introducing additional disorder. Due to large spin-orbit coupling and layer hybridization, tWSe$_2$ is effectively a one-orbital triangular lattice Hubbard model simulator, where a correlated insulating phase has been previously observed at half-filling \cite{Wang2020}. This insulating state can be turned metallic by both electrostatic doping and vertical electric field, giving us sensitive control over a large phase space where the transport properties can be systematically measured. 

We start by describing the boundaries of insulating behaviour as a function of doping and electric field for a particular sample, with a twist angle of 4.2˚. Similar phenomenology is seen for samples at nearby twist angles (see S.I.). The details of our sample fabrication and transport measurements are described in the methods section, and a schematic of our measurement geometry is shown in figs. 1a,b. Shown in fig. 1c is the resistance of the sample measured at low temperature (1.6 K) as a function of density and displacement field. Density is controlled by the average potential applied to the two layers of our bilayer system while displacement field is the potential difference between layers. In our homobilayer tWSe$_2$, displacement field changes the bandwidth of the lowest moiré band as well as influences the on-site Coulomb energy via changing the spatial structure of the Wannier functions \cite{Wang2020}. In order to understand which of the regions in figure 1c corresponds to insulating behaviour, we perform temperature-dependent resistivity measurements along two different lines shown in this plot - the blue line corresponding to a constant density ($\nu=-1$) and the purple line corresponding to varying density while keeping the displacement field around $D\sim 0.2 V/nm$. When the sample displays insulating behaviour, i.e. resistance increasing with decreasing temperature, we can define a gap from the measured activation energy of the resistivity (see S.I. for a typical fit). Shown in fig. 1d and 1e are the measured insulating gaps along the blue and purple dashed lines in figure 1c. Both these plots show that there is a well defined region of insulating behaviour, outside which the resistance continues to remain metallic down to low temperature. Our key finding is that the insulating gaps as a function of both doping and displacement field smoothly approach zero at the metal-insulator boundary, indicating a second order transition. The scenario that emerges from these activation energy measurements is that there is an elliptical region around half-filling that displays insulating behaviour, bounded by a ring of continuous metal-insulator transitions and providing convenient  experimental access to both doping and bandwitdth controlled metal-insulator transitions.

 We now describe the doping-driven metallic state immediately adjacent to the insulator described in fig. 1. In practice, the doping and displacement field are controlled \textit{via} top and bottom gate electrodes separated by hBN insulators from the sample. Shown in fig. 2a is the temperature dependence of the resistivity at a fixed value of top gate voltage over the temperature range 1.5-25 K and doping range $\nu=-0.85$ to $-1.25$. The first derivative of resistivity with temperature ($\partial \rho /\partial T$) from fig. 2a is shown in fig. 2b to better illustrate the temperature dependence. In this plot, insulating regions with a negative $\partial \rho /\partial T$ have been assigned a white color. This allows us to clearly visualize the metal-insulator boundary which forms a dome around $\nu=-1$, extending to a maximum temperature of about 10 K at this top gate voltage. Two doping-driven metal-insulator transitions are seen near $v=-1.1$ and $v=0.9$. Plotted in Fig 2c and 2d are the resistivity curves for different densities in the vicinity of these two transitions. The blue dashed line on these plots shows the decrease in the metal-insulator transition temperature as the transition point is reached on both plots. At the transition itself, the temperature dependence of the resistivity is $T$-linear down to the lowest temperature (1.5 K for this data set, but down to 200 mK in other data sets and across several samples). We see from the color plot in Fig 2b that the $T$-linear behaviour extends up to approximately 20 K at this value of gate voltage, beyond which the temperature dependence becomes sub-linear. This basic phenomenology of T-linear behaviour near the metal-insulator transition is reproduced for several displacement fields for each sample, and across three samples at different twist angles (see SI for an example data set from a different sample).

We have so far described the $T$-linear transport immediately adjacent to the doping-dependent metal-insulator transition at low temperature. We now describe the transport over a wider range of doping. Technically, we cannot access all dopings for all displacement fields due to constraints associated with the maximum gate voltages we can apply before breakdown. To show the wide-range doping dependence, we use a fixed top gate value of $-8 V$ for the same sample as shown in figure 2. At this top gate voltage, the sample displays a smaller gap than shown in figure 2 but is otherwise similar. Shown in fig. 3a is the temperature dependence of the resistivity at this top gate voltage over the wide doping range $\nu=-0.6$ to $-2.51$. Shown in fig. 3c are a series of line traces of resistance versus temperatures at various doping levels. At high doping (well beyond half-filling), the resistivity is fit well by a $T^2$ temperature dependence at low temperature. The region of temperature over which this $T^2$ fit holds is up to about 80 K at a doping of $\nu=-1.6$. As we approach the metal-insulator transition from high doping, the maximum temperature to which the $T^2$ fit holds starts to decrease, and a region of T-linear resistance starts to appear above this temperature. Near the metal-insulator transition, the resistance obeys T-linear behaviour down to the lowest temperatures of our measurement. Upon entering the insulating phase itself, metal-insulator transitions are seen as a function of temperature, and the resistance is T-linear at temperatures just above the insulating phase. Upon reducing the doping further, a second metal-insulator transition is seen, near which the resistivity is again T-linear down to the lowest temperature. Finally, at an even smaller doping of $\nu=-0.84$, $T^2$ resistivity is recovered at low temperature. The overall temperature dependence of the resistivity is summarized in fig. 3b. It shows the presence of two quantum critical fans of T-linear behaviour that project to near the metal insulator transitions on either side of the insulating region, with Fermi-liquid $T^2$ behaviour recovered away from the quantum critical region.
 
In some strongly correlated materials like the cuprate superconductors, the $T$-linear behaviour continues to be observed to anomalously high temperatures, leading to resistances that are well above the Ioffe-Regel limit. In our samples, we do not observe this behaviour. Instead, we observe an equally interesting saturation of the resistance at temperatures of order 200 K, which can be seen in the traces shown in figure 3c. The value of the saturated resistivity is not universal, but displays a weak doping dependence in the range of 3-6 k$\ohm$. In twisted WSe$_2$ at the twist angles of our samples (4-5\degree), theory indicates that the bandwidth is of the order of 50meV. Further, both theory and experiment indicate that there is no true gap between the lowest moiré subband and higher bands. Thus, thermal excitation of carriers to higher bands would seem to be an unlikely explanation for this intriguing phenomenon. 

In Fermi liquid theory, the quadratic parameter $\alpha_Q$ in the relation $\rho (T) = \alpha_QT^2 + \rho_0$ is typically related to the square of the effective Fermi temperature $\alpha_Q\sim T_F^{-2}$. Systems with a quantum critical point exhibit a $1/(n-n_{critical})$ divergence of $\alpha_Q$, indicating the collapse of the Fermi temperature \cite{Qimiao}. To investigate this signature of quantum criticality, we show in the bottom panel of fig. 3c the doping dependence of $\alpha_Q$ for the data set shown in figure 3a. Our data clearly shows that $\alpha_Q$ increases by over an order of magnitude near the quantum critical points $\nu=-1.24$ and $\nu=-0.94$. In contrast, we find very little dependence of $\alpha_Q$ in the metallic region beyond $\nu=-2$. Along with $\alpha_Q$, we plot in fig. 3a the parameter $\alpha_L$ from linear fits to resistivity of the form $\rho (T) = \alpha_LT + \rho_0$. This parameter shows a clear maximum at the quantum critical points, where T-linear behaviour is observed to lowest temperature. From Drude model considerations, we can obtain an estimate of the scattering time $\tau$ of the quasiparticles in the system. Similarly to Cao et al. \cite{Cao2020}, we can assign a dimensionless parameter $C$ such that $\tau=\frac{\hbar}{Ck_BT}$ for which, if $C\rightarrow1$, the system nears the Planckian dissipation regime. For tWSe$_2$, the effective mass is $\sim 0.4m_e$ \cite{Kormanyos} and $n\sim 0.5\times10^{12}cm^{-2}$, then $C=\frac{\hbar e^2 n}{k_B m^*}\alpha_L\sim 0.027\alpha_L$. In our case, $C$ ranges from 1 to 10. 

Along with the T-linear behaviour generically observed at quantum critical points, unconventional superconductors such as the cuprates and pnictides exhibit B-linear magnetoresistance above the putative quantum critical point \cite{Bruin,Gallo,Analytis}. In these materials, superconductivity obscures the low-field, low temperature limit. Reliable magnetoresistance data requires multiple good contacts to the sample to separate the longitudinal and Hall resistance cleanly, which is a challenge for tWSe$_2$ samples, especially at low doping. However, we are able to obtain reliable data at fillings above half filling at selected displacement fields. Shown in fig. 4a and b is a set of curves of the longitudinal and Hall magnetoresistance for a sample with a twist angle of 4.5˚ at a top gate voltage of -19 V, for which the maximum insulating gap at half filling is $\sim 0.7$meV. Similar data is obtained on other samples (see S.I. for another example). The data are shown for doping values ranging from $\nu=-1.2$ (red curves) to $\nu=-1.02$ (blue curves). The low field longitudinal magnetoresistance shows a striking evolution from a weak $B^2$ dependence at $\nu=-1.2$ to a strong B-linear dependence near the metal-insulator transition. The entire set of data at low field (below 1 T) is well described by an ansatz previously proposed for the cuprates and pnictides $\rho(B) = \sqrt{\gamma+\beta B^2}$ \cite{Analytis}. The extracted values of $\beta$ as a function of $\nu$ is shown in fig. 4c, clearly showing a large increase on approaching the insulating phase, corresponding to a crossover from $B^2$ to $B$-linear behaviour. In the cuprates and pnictides, it has been phenomenologically observed that the slope of the B-linear magnetoresistance and the linear-in-T dependent resistance at zero field are comparable by converting magnetic field to temperature via the relationship $g\mu_BB=k_BT$. Such a conversion in our case would imply that $\sqrt{\beta}=\frac{g\mu_B}{k_B}$. Indeed, assuming $g^*\sim 25$ near half-filling \cite{Tutuc_g} and converting the values of $\sqrt{\beta} = 600-800$ $\Omega/T$ using this relationship gives $\alpha_L=36-48$, close to the experimentally observed values of 46 $\Omega/K$ at zero field for this particular sample and displacement field. In analogy with cuprates and pnictides, the implication is that magnetic field and temperature both play equivalent roles in determining the scattering rate in the quantum critical region.

So far, we have discussed the phase diagram at a fixed top gate voltage and described transport near the doping-driven quantum critical points. We now describe the phase diagram as a function of displacement field at half filling. Shown in  fig. 5a are a series of line traces at displacement fields going across the metal insulator boundary as described by the green line in fig. 1c. At the lowest displacement field among these traces (D=0.12 V/nm), the resistance shows activated behaviour with a gap of $\sim 2.4$meV. On increasing the displacement field, the gap decreases continuously until it goes to zero. At a displacement field of $D_{critical} = 0.33 V/nm$, the resistivity is found to be T-linear to the lowest temperature (see the inset to fig. 5a). At even higher values of displacement field, $T^2$ behaviour is recovered in the low temperature resistivity (also in the inset to fig. 5a). 

Hartree-Fock calculations show (see SI)  that  for band dispersions believed relevant to t-WSe$_2$, and for interaction strengths in a physically reasonable intermediate coupling range, a metal-insulator-metal sequence of  transitions occurs as the displacement field is varied. Theoretical evidence \cite{Motrunich05,Sahebsara08} is that the insulating phase is non-magnetic, does not break a spatial symmetry, and is most probably a spin liquid of some type, although the precise identification  remains controversial. Very recent results suggest that the transition is second order \cite{Moore20}. The quantum critical scaling we observe in the vicinity of the transitions suggests that the insulating phase is characterized by an order parameter whose fluctuations go soft at the transition and can scatter electrons.  

Having established that both the doping and displacement field metal-insulator transitions host quantum critical points, we proceed to investigate the metallic phase in the entire $n-D$ phase diagram \textit{via} temperature dependent resistance measurements. Shown in fig. 5b is a color map of insulating gaps (in shades of red) observed in the $n-D$ phase diagram. The color scale indicates the gap at each value of $n$ and $D$, established by activation energy fits. The light pink region indicates metallic behaviour to lowest measured temperature. T-linear behaviour down to base temperature is observed near the entire metal-insulator boundary. Shown in fig. 5c is a color plot of the slope of the T-linear resistivity (in regions where temperature dependence is linear) for the same sample. The data clearly show that both the slope of the T-linear behaviour and the magnitude of the insulating gap show a similar evolution with displacement field. This strong correlation between the T-linear behaviour and the magnitude of the correlated gap, as well as the presence of $T^2$ resistivity far from the quantum critical points, indicate that in this temperature regime the resistance is dominated by strong electronic correlations rather than by electron-phonon scattering. In this regard, the situation is different from twisted bilayer graphene \cite{Cao2020, Polshyn2019}, where a debate exists about the role of electron-phonon scattering in the observed T-linear resistivity \cite{Polshyn2019, DasSarma_graphene, Adam}.

In the metallic phase, the resistance drops with decreasing temperature down to the lowest temperatures of our measurement (typically between 200 mK and 1.6 K). The resistivity can be extrapolated to zero temperature to give a residual resistivity $\rho_0$ that is small when deep in the metallic phase (see Fig. 5e). On the other hand, in the insulating state the resistivity increases rapidly as T is decreased below the metal-insulator transition, but for higher temperatures the resistivity is approximately linear in $T$. A ``residual resistivity" may be defined by extrapolating this linear behaviour to $T=0$ as shown in figure 5d. A very strong association of this residual resistivity with the strength of the insulating behaviour is evident from Fig. 5e and this association is quantified in Fig. 5f.

The behaviour of the residual resistivity shown in figure 5f has two interesting features. In pure conventional metals, the resistance is on general grounds expected to vanish at $T=0$; a residual resistivity arises from impurities and study of the residual resistivity near a quantum critical point has been complicated by the need to fabricate different samples, with potentially different impurity concentrations. A unique advantage of moiré materials is that the transition is driven by doping or displacement fields within a single sample and therefore at fixed impurity concentration. From our results it is evident that the residual resistivity is small and weakly doping-dependent in the metallic phase, but rises quickly (by as much as $\sim 10$k$\Omega$) when we enter the insulating phase either through bandwidth control or doping control. Furthermore, with $\nu$ fixed at -1, we find that the gap fully opens when $\rho_0$ is between 2 and 3k$\Omega$. This behaviour must be understood in terms of the interplay of impurity scattering and scattering from fluctuations of the order parameter of the insulating state, which become singular at the metal-insulator transition. Indeed, theories that predict spin liquid behaviour in continuous Mott transitions do predict such a rapid rise in the residual resistivity at the critical point \cite{Senthil}. The second interesting behaviour observed in figure 5f is that the residual resistivity as defined from the extrapolation to $T=0$ of the $T$-linear part of $\rho$ continues to stay large as we cross from the metallic phase into the insulator. This behaviour could be related to the presence of fluctuations of the insulating order parameter above the metal insulator transition that give rise to additional scattering. 

In conclusion, the transport data presented here provide compelling evidence of a continuous, quantum critical transition from a metallic to an insulating state, as doping and bandwidth are varied in a system of interacting fermions on a triangular lattice with a band filling near one electron per lattice site. The experimental platform provides access to the temperature, doping and bandwidth controlled transition within one sample, removing the ambiguities associated with the need to fabricate a range of samples that have plagued previous work. The reentrant nature of the transition with displacement field establishes that the interaction strength is moderate, of the order of the bandwidth and prior theoretical evidence strongly suggests that the moderate correlation insulating phase is exotic, most likely a spin liquid. Furthermore, theoretical predictions on the doping-tuned metal-insulator transition indicate a wealth of interaction-driven metallic phases in this particular system \cite{Pan2}. Our results strongly suggest that, whatever its origin, the insulating phase is characterized by fluctuations that can scatter electrons, become soft at the transition, and have a nontrivial effect at higher temperature above the scale at which the insulating gap opens. Our work opens new directions in the study both of spin liquids and of correlation-driven insulators more generally.

\newpage

\bigskip

\bibliographystyle{apsrev4-1}
\bibliography{refs.bib}

\bigskip
\section*{Acknowledgments}
 The authors thank Jennifer Cano, Antoine Georges, Chandra Varma, Tomo Uemura, T. Senthil, Ivan Bozovic, Matt Yankowitz, Sankar Das Sarma, Qimiao Si, Alex Wietek and Jie Wang for fruitful discussions. AG, AJM and LW are supported in  part by Programmable Quantum Materials, an Energy Frontier Research Center funded by the U.S. Department of Energy (DOE), Office of Science, Basic Energy Sciences (BES), under award DE-SC0019443. EMS, DAR, BK, JH and CRD are supported by the Columbia MRSEC on Precision-Assembled Quantum Materials (PAQM) - DMR-2011738. ANP is supported by the Air Force Office of Scientific Research via grant FA9550-16-1-0601. The Flatiron Institute is a division of the Simons Foundation. 

\bigskip
\section{Author Contributions}
A.G., L.W., E.-M.S. and D.A.R. fabricated the samples. A.G., L.W. and E.-M.S. performed the transport measurements. A.G. and G.S.S.G.P. analysed the data. D.A.R., B.K. and J.H. grew the WSe$_2$ crystals. K.W. and T.T. grew the hBN crystals. J. W. and A.J.M. supervised the theoretical aspects of this work. A.G., A.N.P., C.R.D. and A.J.M wrote the manuscript with input from all the authors.

\bigskip
\section{Competing interest declaration}
The authors declare no competing interest.

\section{Methods}

Sample preparation is discussed in detail in Wang et al. \cite{Wang2020}. Twisted WSe$_2$ samples were prepared using the dry stamp and 'tear-and-stack' technique \cite{wang_contact,tutuc_theta}. tWSe$_2$ is first picked up using the top hBN flake and then placed on top of pre-patterned Cr/Pt contacts \cite{tutuc_contact} on the bottom hBN. A combination of metal and/or graphite is used as top and bottom gates as shown in fig. 1a.

Doping and displacement field values are obtained from linear extrapolations of Landau fans projecting from the band edge and full-filling of the moiré unit cell. We find these Landau levels to be two-fold degenerate. Then, the capacitance for the top and bottom gates are obtained as well as the full-filling density (or conversely, the twist angle). By assuming a linear gate response of the system, the number of holes per moiré unit cell can be obtained across all dopings \cite{Wang2020}.

\section{Data availability}
The data that support the findings of this study are available from the corresponding authors upon request.

\newpage

\begin{figure*}[t]
\begin{center}
\includegraphics[width=\linewidth]{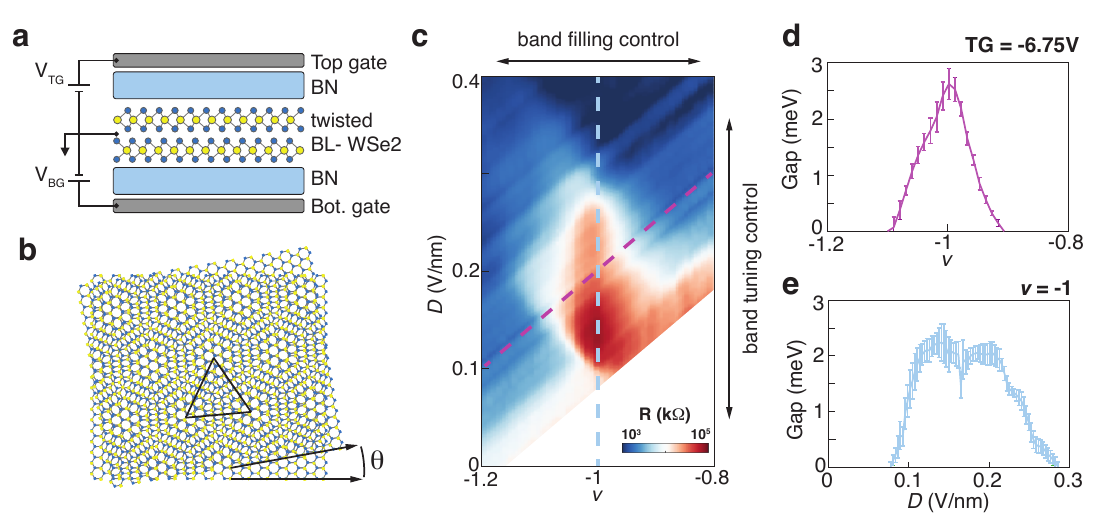}
\caption{   \textbf{Continuous metal-insulator transition in twisted WSe$_2$ }\\
            \textbf{a} Illustration of a dual--gated twisted--bilayer WSe$_2$ device.  Biasing the top gate ($V_{TG}$) and bottom gate ($V_{BG}$) allows independent control of the transverse displacement field and carrier density. \textbf{b} Cartoon representation of bilayer WSe$_2$, twisted to angle $\theta$.  The resulting moiré pattern acts like a superlattice potential with triangular symmetry. \textbf{c} Low temperature resistance ($T=1.5 K$) plotted versus displacement field, $D$, and band filling, $\nu$, for a sample with twist angle $\theta=4.2$\degree. The band filling is defined in units of electrons per unit cell of the moiré superlattice.  Varying the displacement field allows in-situ bandwidth tuning at fixed filling \cite{Wang2020}.  This makes it possible to map the metal-insulator phase diagram through both band filling and band tuning control mechanisms \cite{Imada}.  \textbf{d,e}. Plot of the insulating gap, measured along the lines of corresponding color shown in \textbf{c}.  For both sets of measurements, gaps are extracted from Arrhenius fits to the resistivity. At the metal-insulator boundary, it is observed that the insulating gap goes to zero continuously.
}

\label{fig:fig1}
\end{center}
\end{figure*}

\newpage

\begin{figure*}[t]
\begin{center}
\includegraphics[width=0.9\linewidth]{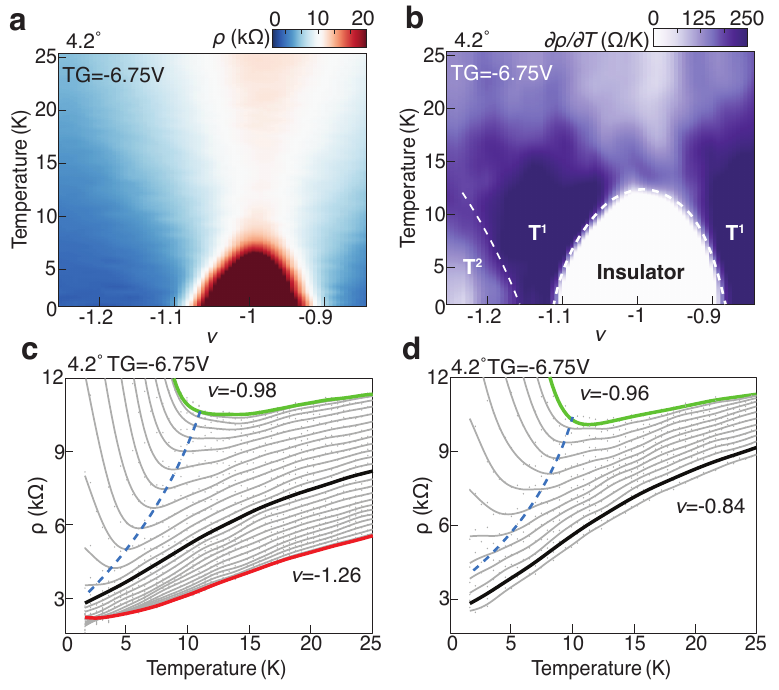}
\caption{
    \textbf{Doping-driven metal-insulator transition in twisted WSe$_2$ } \\
        \textbf{a,} color plot of resistivity versus temperature and doping for a 4.2˚ device at a fixed top gate value of -6.75V.  
        \textbf{b,} First derivative in temperature of the resistivity from \textbf{a}. All values below zero are set to zero (white color). At the boundaries of the insulating region, $T$-linear resistivity is observed at low temperature.   
        \textbf{c,d} Line plots of the resistivity for doping ranges near the zero-temperature metal-insulator transitions. The green highlighted curves display insulating behaviour, with a metal-insulator transition defined by the blue dashed lines. At the metal-insulator boundary, the low temperature transport is $T$-linear (black curves). Further into the metal, the resistance starts to develop $T^2$ behaviour at low temperature (red curve in panel \textbf{c}).
}

\label{fig:fig2}
\end{center}
\end{figure*}

\newpage

\begin{figure*}[t]
\begin{center}
\includegraphics[width=1\linewidth]{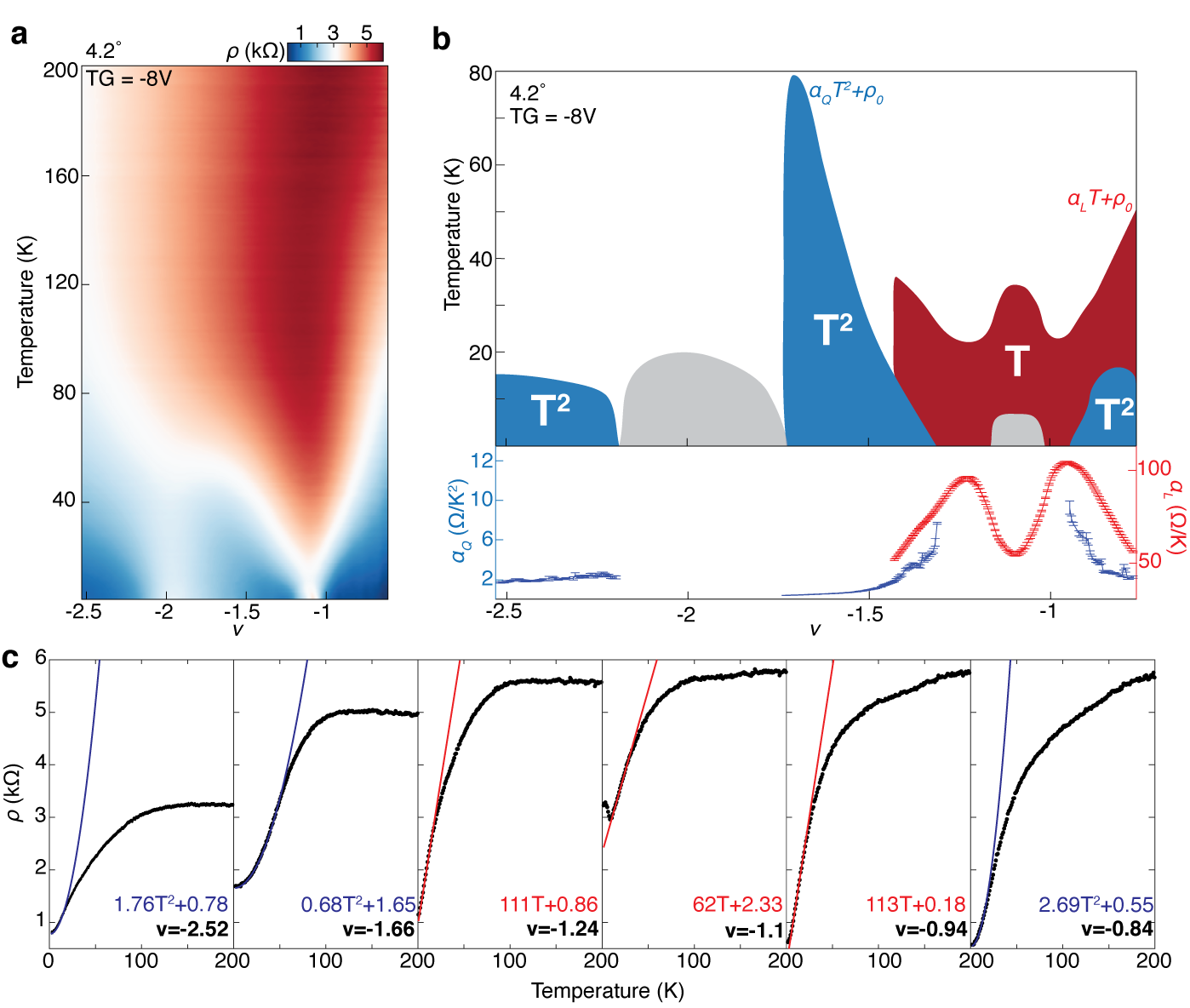}
\caption{
    \textbf{Quantum critical fan in twisted WSe$_2$} \\
    \textbf{a,} Color plot of resistivity versus temperature and doping for a 4.2˚ device at a fixed top gate value of -8V.  
    \textbf{b,} The upper panel shows a summary of the different temperature dependences observed in this measurement. Blue color denotes $T^2$ behaviour, red indicates $T$-linear behaviour and grey indicates insulating behaviour. Two quantum critical fans are seen to emerge on either side of the insulating state near half-filling, but no such behaviour is observed near the trivial insulator at full filling. The bottom panel shows the evolution of fit parameters with doping. The magnitude of the $T^2$ coefficient $\alpha_Q$ is plotted in shades of blue (left y-axis), and the magnitude of the $T$-linear  coefficient $\alpha_L$ is plotted in red (right y-axis). $\alpha_Q$ shows a critical enhancement by over an order of magnitude as the quantum critical point is approached from the metallic phase. $\alpha_L$ a maximum at the quantum critical points where linear-in-T resistivity is observed down to lowest measured temperature.  
    \textbf{c,} Plots of resistivity (black) versus temperature as well as quadratic (blue) and linear (red) fits for different dopings. At temperatures above 100 K the resistivity starts to saturate to a doping dependent, but temperature-independent, behaviour. 
}

\label{fig:fig3}
\end{center}
\end{figure*}

\newpage

\begin{figure*}[t]
\begin{center}
\includegraphics[width=1\linewidth]{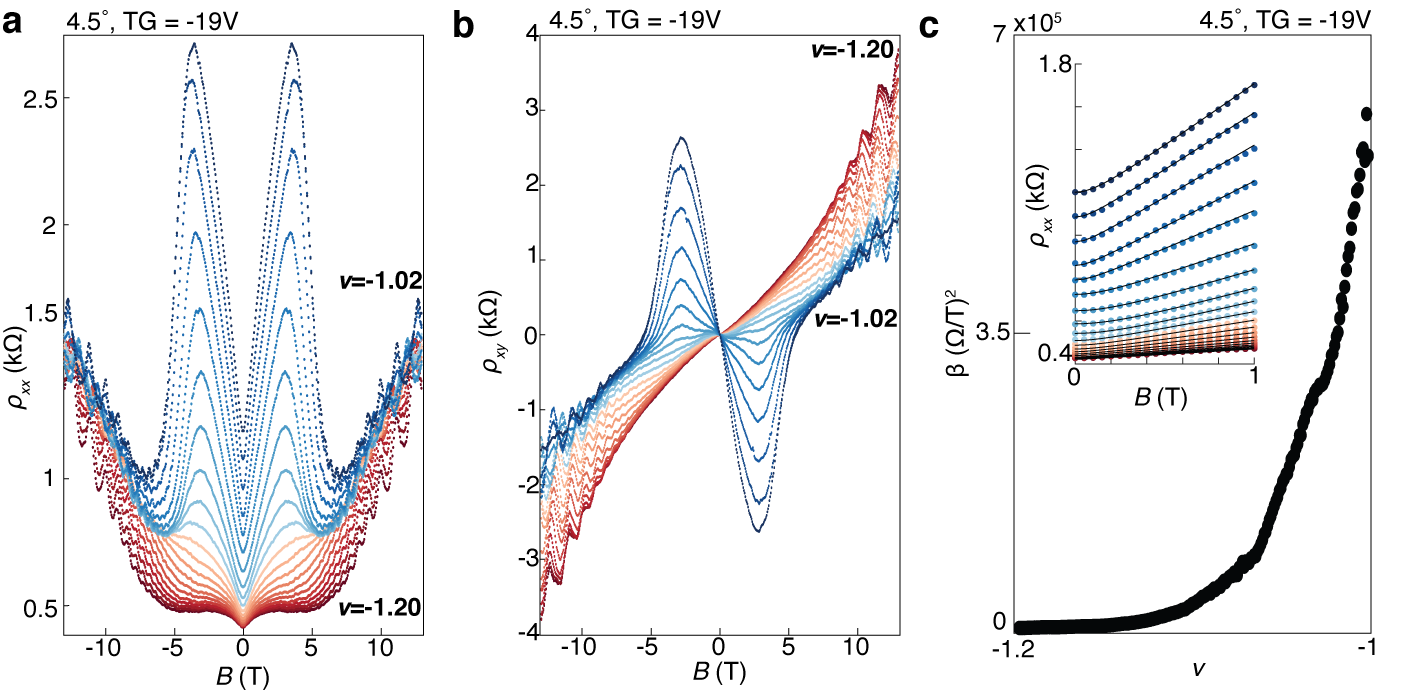}
\caption{
    \textbf{Anomalous magnetotransport} 
    \textbf{a,} Doping dependence of the longitudinal magnetoresistance for a sample at a twist angle of 4.5˚, in the metallic region in the vicinity of the quantum critical point. At low fields (0-1T) the resistivity smoothly evolves in doping from a $B^2$ form deep in the metallic phase (lowest curve) to  $B$-linear behaviour in the vicinity of the quantum critical point (upper curves).
    \textbf{b,} Doping dependence of the Hall resistance over the same range shown in \textbf{a}. The Hall resistivity shows dramatic changes associated with the collapse of the Fermi surface on approaching the quantum critical point.
    \textbf{c,} Fit results of longitudinal resistivity in \textbf{a} from 0 to 1T to $\sqrt{\gamma +\beta B^2}$ with a confidence interval greater than 95\%. We observe a critical increase in the parameter $\beta$ as it nears half-filling, corresponding to a transition from $B^2$ to $B$-linear behaviour. Sample fits are displayed in the inset.
}

\label{fig:figS1}
\end{center}
\end{figure*}

\newpage

\begin{figure*}[t]
\begin{center}
\includegraphics[width=1\linewidth]{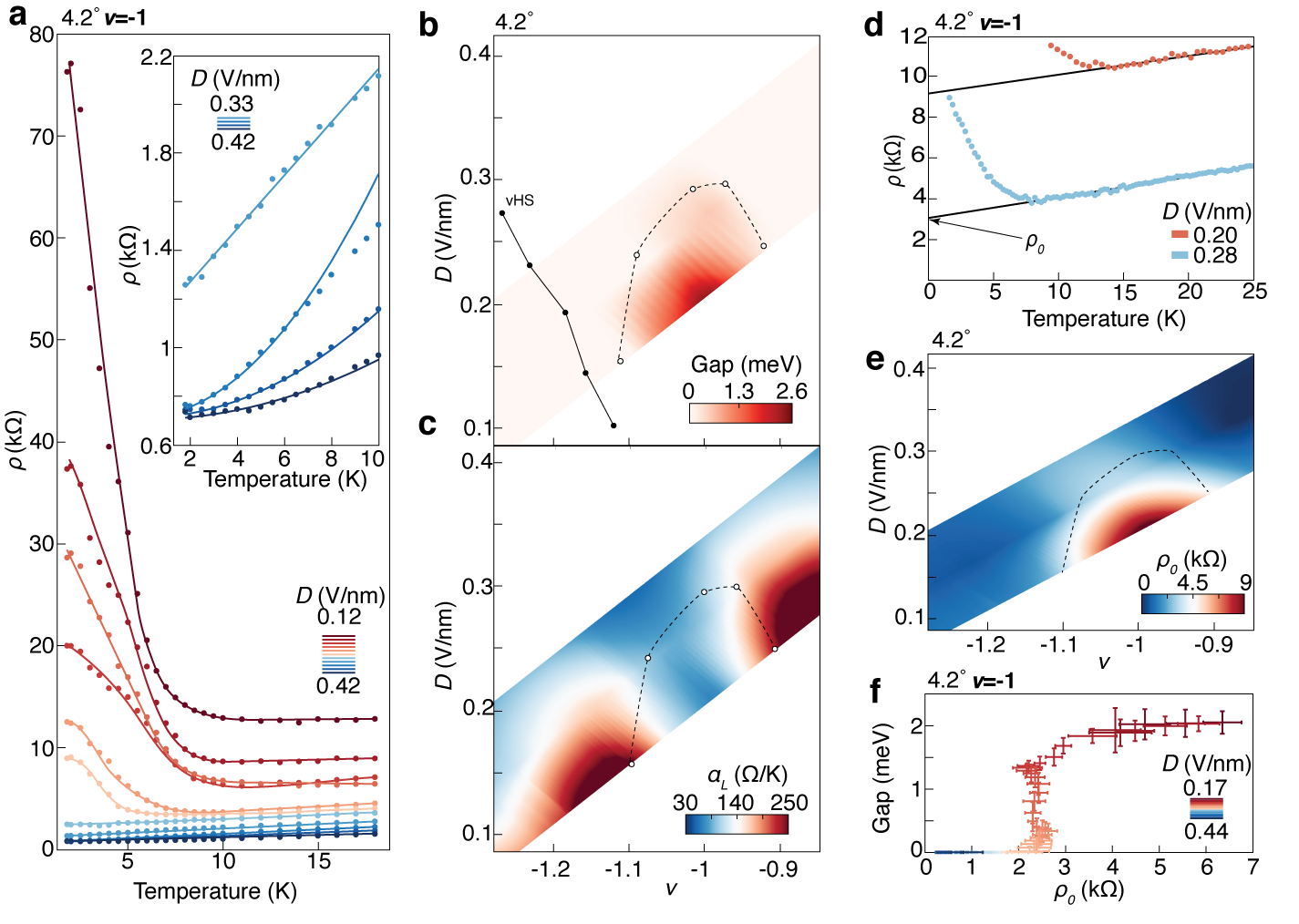}
\caption{
    \textbf{Bandwidth-driven quantum criticality}  \\
    \textbf{a,} Resistivity versus temperature at half-filling for a range of displacement fields over which a metal-insulator transition is observed at half-filling. Inset: Detailed dependence at low temperature in the metallic phase near the quantum critical point. At the quantum critical point (D=0.33 V/nm, top curve) T-linear behaviour is seen down to low temperature, while at higher displacement fields $T^2$ behaviour is observed.
    \textbf{b,} Color plot of insulating gaps (red colors) as a function of density and displacement field. The boundary of the insulating region is shown by the dashed line. The position of the van Hove singularity obtained from Hall measurements is also plotted.
    \textbf{c,} Linear coefficient $\alpha_L$ versus displacement field and doping over the same range as \textbf{b}. Both $\alpha_L$ and the magnitude of the insulating gap itself are controlled by the displacement field, indicating a common origin. The boundary of the gapped region in \textbf{b} is denoted by the dashed line in \textbf{c}.
    \textbf{d,} $\rho_0$ defined from extrapolation to $T=0$ of high T  region of ``metallic"  resistivity for insulating regions of the phase diagram. 
    \textbf{e,} The evolution of $\rho_0$ for doping and displacement field. $\rho_0$ is strongly correlated with the magnitude of the insulating gap shown in \textbf{b}.
    \textbf{f,} Gap versus $\rho_0$ at half-filling. The insulating regions display a $\rho_0$ of at least 2.5 k$\Omega$ while in the metallic phase $\rho_0$ quickly approaches zero. 
   }

\label{fig:fig4}
\end{center}
\end{figure*}

\end{document}